\documentstyle[preprint,revtex,eqsecnum]{aps}
\newcommand{\beq}{\begin{equation}}
\newcommand{\eeq}{\end{equation}}
\begin{document}
\draft
\preprint{OS-GE-27-92}
\preprint{DPNU-92-48}
\preprint{Feb. 1993}
\begin{title}
   Unified Description of Hadrons\\
   and Heavy Hadron Decays
\end{title}
\author{Noriaki Kitazawa\cite{Kitazawa}}
\begin{instit}
   Department of Physics, Nagoya University\\
   Negoya 464-01, Japan
\end{instit}
\author{Takeshi Kurimoto\cite{Kurimoto}}
\begin{instit}
Institute of Physics, College of General Education,
Osaka University\\
Toyonaka 560, Osaka, Japan
\end{instit}
\begin{abstract}
We construct an effective Lagrangian which describes interactions
of heavy and light hadrons utilizing the chiral flavor symmetry for
light quarks and heavy quark symmetry. For both light and heavy
sector we include pseudo scalars, vectors and baryons
in the Lagrangian. Heavy hadron decays are discussed as application
of our formalism. The $D_s$ decay constant and the
coupling constant among heavy meson, heavy vector meson and light
meson are fitted from the experimental data
of $D^0 \rightarrow K^- e^+\nu_e$ decay.
We also point out the possibility of model independent determination
of $|V_{ub}|$ from $\Lambda_b \rightarrow p l \bar \nu$ decay.
\end{abstract}
\section{Introduction}
\label{sec:intro}
Investigation of physics concerning hadrons has been suffering from
the uncertainty in evaluating the strong interaction effect.
Various approaches of estimating non-perturbative strong interaction
effect have been made so far, but still we need more effort to make
our predictions precise. The approaches may be divided into two
categories. One is direct calculation of physical
quantities, for example, lattice QCD calculation, potential model
calculation and so on. The other is symmetry approach where one
quantity is related to another by using the symmetry of hadrons.
Predictions are made from a few experimental or theoretically
estimated values. A merit of the latter approach
is that the predictions are independent of the underlying dynamics
of the hadron as far as the symmetry is good. So the hadron model
dependence can be removed.

The chiral $SU(3)_R \times SU(3)_L$ symmetry has been frequently
used in analyzing the physics of light hadrons
composed of $u$, $d$ and/or $s$ quarks.
Chiral Lagrangian method has been developed
as a powerful tool to utilize this symmetry\cite{Georgi1}.
A Lagrangian of hadron (not quark) fields is constructed
to realize the chiral symmetry which spontaneously breaks down
to $SU(3)_V$. The form of hadron interaction is determined by
the symmetry. The values of coupling constants are determined by
fitting the predictions with the experimental data.

Recently Isgur and Wise proposed the usefulness of
another kind of hadron symmetry which is
available for hadrons containing a heavy quark like $c$ and $b$
\cite{WIsgur,HQrev}.
They showed that there exist two kinds of symmetry in the limit
of infinite quark mass. One is to rotate flavors among heavy quarks,
and the other is to rotate spin of quark separately
in each heavy flavor.

Burdman and Donoghue\cite{BurDon},
Wise\cite{Wise},  Yan et.al. \cite{Yans} and
Casalbuoni et.al.\cite{Casa} combined the chiral symmetry of
light quarks and heavy quark symmetry to build an effective
Lagrangian of heavy hadrons and light pseudoscalar mesons.
Wise used an elegant formalism to express both heavy pseudoscalar
and vector mesons simultaneously\cite{Wise}.
Cho extended this formalism to include heavy baryons\cite{Cho}.
Casalbuoni et. al. included light vector mesons as the gauge bosons
of hidden local symmetry.\cite{Casa}.In this work we extend these
works to describe the interaction among hadrons
from pions, light vector mesons to heavy baryons including light
baryons with the formalism used in the Wise's paper.
An unified description of hadrons is given in terms of an effective
Lagrangian by utilizing the chiral symmetry and heavy quark symmetry.
Then we apply our effective Lagrangian to some heavy hadron decays.
$D^0 \rightarrow K^- e^+ \nu _e$ decay is analyzed to determine
the $D_s$ decay constant and the coupling constant among heavy
pseudo scalar meson, heavy vector meson
and light pseudoscalar meson.
$\Lambda_b \rightarrow p l \bar\nu$ decay are investigated as an
interesting mode of determing $|V_{ub}|$. We point out the
possibility of model independent determination of $|V_{ub}|$ in
this decay mode.

The rest of this paper is organized as follows.
In Sec.~\ref{sec:sym} we review the
symmetries and transformation properties of hadrons.
In Sec.~\ref{sec:lag} the Lagrangian of strong interaction part is
constructed. In Sec.~\ref{sec:weak} weak currents are given in
terms of hadron fields to deal with weak interaction.
Sec.~\ref{sec:apl} is devoted to application of our formalism to
weak decay of heavy hadrons.
Sec.~\ref{sec:conc} is concluding remarks.
\section{Symmetries and transformation properties of hadrons}
\label{sec:sym}
\subsection{Chiral flavor symmetry of light sector
	and its non-linear realization}

If the masses of N different flavored quarks are zero,
there is a global symmetry of
rotating the flavor of left handed and right handed quarks
separately, i.e.  $U(N)_R \times U(N)_L$ symmetry.
This symmetry spontaneously breaks down to $U(N)_V$ due to
non-zero $\langle\bar qq \rangle$ condensation.
Although quarks are not massless, we have $U(3)_R \times U(3)_L$
symmetry in a good approximation because the current masses
of $u$, $d$ and $s$ quarks are much smaller than the chiral
symmetry breaking scale. The light pseudoscalar mesons can be
regarded as the Nambu-Goldstone bosons, while the light vector
mesons like $\rho$ as the gauge bosons of hidden local $U(3)_V$
symmetry. These hadrons are described by using the non-linear
realization of $U(3)_R \times U(3)_L/U(3)_V$ symmetry as proposed by
Bando, Kugo, Uehara, Yamawaki and Yanagida\cite{BKY}.
The light pseudoscalar mesons are expressed as coset
space coordinates of $U(3)_R \times U(3)_L/U(3)_V$;
\beq
\xi\equiv e^{i\pi /f},\qquad
     \pi\equiv\frac{1}{\sqrt{2}}\left(
            \begin{array}{ccc}
                   (\pi^0+\eta_1)/\sqrt{2} & \pi^+ & K^+ \\
                    \pi^-      & (-\pi^0+\eta_1)/\sqrt{2} & K^0 \\
                    K^-  & \bar K^0 & \eta_3
            \end{array} \right),
\eeq
where $\eta_1\sim (u\bar u + d\bar d)/\sqrt{2}$ and
$\eta_3\sim s\bar s$.

The light vector mesons are expressed as
\beq
V\equiv\frac{g_V}{\sqrt{2}}\left(
            \begin{array}{ccc}
                (\rho^0+\omega)/\sqrt{2} & \rho^+ & K^{*+} \\
                 \rho^-      & (-\rho^0+\omega)/\sqrt{2} & K^{*0} \\
                 K^{*-}  & \bar K^{*0} & \phi
            \end{array} \right).
\eeq
They transform under global $U(3)_R \times U(3)_L$ symmetry and
hidden local $U(3)_V$ symmetry as
follows,
\beq
\xi \rightarrow z_L \xi {z_V}^{\dag}=z_V \xi {z_R}^{\dag},
\qquad  V\rightarrow z_V V {z_V}^{\dag}-i \partial z_V{z_V}^{\dag},
\eeq
where $z_R$,  $z_L$ and  $z_V$ are the elements of
$U(3)_R$, $U(3)_L$ and $U(3)_V$, respectively. It is convenient
to define the following 1-forms,
\beq
	\alpha_\perp\equiv
	\frac{i}{2}(\xi\partial\xi^{\dag}-\xi^{\dag}\partial\xi),
\qquad
	\alpha_\parallel\equiv
	\frac{i}{2}(\xi\partial\xi^{\dag}+\xi^{\dag}\partial\xi).
\eeq
The 1-form $\alpha_\perp$ transforms homogeneously under
the $SU(3)_V$, while $\alpha_\parallel$ in the same way
as the gauge field $V$.

The light baryon octet are introduced
as $(3,\ 3^*)$ representation of $SU(3)_V$\cite{Georgi1}.
\beq
\psi\equiv\left(
   \begin{array}{ccc}
     \frac{\Sigma^0}{\sqrt{2}}+\frac{\Lambda}{\sqrt{6}} &
        \Sigma^+ & p \\
     \Sigma^- & -\frac{\Sigma^0}{\sqrt{2}}+
     	\frac{\Lambda}{\sqrt{6}} & n \\
     \Xi^-  & \Xi^0 & \frac{-2\Lambda}{\sqrt{6}}
            \end{array} \right).
\eeq
It transforms as
\beq
\psi  \rightarrow  z_V\psi {z_V}^{\dag}.
\eeq
\subsection{Flavor and spin symmetry of heavy sector}

Suppose the emission of a soft gluon (soft pion) from a heavy
quark (hadron) of momentum $p$. The momentum conservation
gives $p=p'+k$,where $p'$ and $k$ are the momentum of heavy field
and gluon (pion), respectively, after the emission. If the mass of
the heavy field, $M$, is much larger than the scale of $k$,
we have $p/M=p'/M$ neglecting the soft momentum $k$. That is,
the velocity of the heavy field conserves during the
interaction. We can use the velocity instead of the momentum to
characterize the heavy field\cite{Georgi2}. The QCD Lagrangian
reduces in the limit of infinite quark mass to
\beq
{\cal{L}}^H_{QCD} = i \bar h_{Qv} v\cdot D h_{Qv},
\eeq
where $h_{Qv}$ is the effective heavy quark field of
velocity $v$\cite{Georgi2}.
This Lagrangian has the symmetry of rotating heavy flavor and the
symmetry under the rotation of spin in each heavy quark flavor.
We require the effective heavy meson fields
to transform in the same way as the interpolating fields,
\beq
P\sim \bar q i\gamma_5 h_{Qv}, \qquad
P^*_\mu \sim \bar q (\gamma_\mu-v_\mu) h_{Qv}, \label{intp}
\eeq
where the combination $\gamma_\mu-v_\mu$ in the vector meson
interpolating field appears because $v^\mu P^*_\mu=0$ for the
effective field. These effective heavy meson fields are represented
in a unified fashion by introducing the following fields\cite{Wise},
\beq
H \equiv \frac{(1+v\cdot\gamma)}{2} \left(
          P^* \cdot \gamma - P \gamma_5 \right).
\eeq
where
\beq
P^{(*)} = \left( \begin{array}{c}
                       P_D^{(*)}\\
                       P_B^{(*)}
                  \end{array} \right),
\eeq
and the normalization of these boson fields differs from the usual
one by $\sqrt{2M_{D,B}}$. The dimension of these fields is 3/2.
This effective field $H$ transforms as follows under the hidden
local $U(3)_V$, heavy $U(2)$ flavor symmetries and the spin symmetry;
\beq
  H  \rightarrow \left\{
    \begin{array}{ll}
    	H{z_V}^{\dag}& (U(3)_V)\\
    	z_H^v H& (U(2)_{heavy}) \\
    	S^v H =
    	\left(\begin{array}{cc}
    	S_c^v & 0 \\
    	0 & S_b^v
    	\end{array}\right)
    	H & {\rm (spin\ symmetry)}\ ,
    \end{array} \right. \label{tran}
\eeq
where $z_H^v$ and  $S_{c,b}^v$ are the elements of heavy
flavor $SU(2)$ rotation
and heavy spin rotation, respectively, for hadrons of velocity $v$.
The explicit form of the spin rotation is given as
\beq
 S^v= \exp [i\theta_a S^v_a] ,
 \eeq
where $S^v_a=i\epsilon_{abc}[e_b\cdot \gamma, e_c\cdot \gamma]$ with
$e_a$ being a spacelike unit vector which satisfy
$e_a\cdot e_b=-\delta_{ab}$ and $v\cdot e_a=0$.
This spin transformation rotates the pseudoscalar and
the vector effective fields into one another in the same way
as the interpolating fields given in the expression (\ref{intp}).

We consider here the heavy baryons composed of one heavy quark
and rest light degrees
of freedom. The spin degree of freedom of heavy quark is decoupled
from other system
due to the heavy quark spin symmetry. We can classify the
heavy baryons by the total angular
momentum, $J_l$, of light degrees of freedom\cite{IW2}.
For $J_l=0$ we have
$(2,~\bar 3)$ multiplet of $SU(2)_{heavy}\times SU(3)_V$.
It is expressed as
\beq
T_v = \frac{(1+v\cdot\gamma)}{2}
           \left( \begin{array}{ccc}
                       \Xi^0_c & \ -\Xi^+_c &\ \Lambda_c^+ \\
                       \Xi^-_b & \ -\Xi^0_b &\ \Lambda_b
                    \end{array} \right),
\eeq
which transforms in the same way as $H$. Other $J_l$ states can be
introduced as shown in Yan et.al.\cite{Yans} and Cho\cite{Cho},
but we do not consider them here.

The transformation property of hadrons given so far is summarized
in Table \ref{table1}.
\section{Strong interaction Lagrangian}
\label{sec:lag}
\subsection{Lowest order Lagrangian}

Given the transformation properties of hadrons under the symmetries,
the Lagrangian of the strong interaction part can be constructed
by making invariant under the hadron symmetry including $C$ and $P$.
The lowest order Lagrangian in derivative on light hadrons, light
quark mass and inverse power of heavy quark mass are given below.
\begin{eqnarray}
\cal{L} &=& {\cal{L}}_{\xi,V} + {\cal{L}}_{H} + {\cal{L}}_{\psi}
          + {\cal{L}}_{T}, \\
{\cal{L}}_{\xi,V} &=& -\frac{1}{2g_V^2}{\rm tr}(F_V^2)
                 + f^2 {\rm tr}(\alpha_\perp^2)
                 +\frac{m_V^2}{g_V^2}
                     {\rm tr}[(\alpha_\parallel -V)^2], \\
{\cal{L}}_{H} &=& \frac{1}{2} {\rm Tr}[-i\bar H v\cdot\partial H
                 +\bar H H v \cdot \{r\alpha_\parallel+
                                     (1-r)V\}
                 +\lambda \bar H H \gamma\gamma_5 \cdot \alpha_\perp
                  ], \\
{\cal{L}}_{\psi} &=& {\rm tr} \left\{
       \bar\psi i \gamma\cdot\partial\psi + \bar\psi\gamma\cdot
        [s\alpha_\parallel+(1-s)V, \psi] \right\}
        -\mu {\rm tr}(\bar\psi\psi)  \nonumber \\
         & & +\omega_1 {\rm tr}
         (\bar\psi\gamma\gamma_5\cdot\alpha_\perp\psi)
             +\omega_2 {\rm tr}
         (\bar\psi\gamma\gamma_5\psi\cdot\alpha_\perp),\\
{\cal{L}}_{T} &=& i{\rm tr}[\bar T v\cdot \partial T ]
                 -{\rm tr}[\bar T T v\cdot
                     \{t\alpha_\parallel+(1-t)V\}],
\end{eqnarray}
where the trace with capital letter includes trace in Lorentz
spinor indices also, while small letters in flavor indices only.
The parameters $f$, $g_V$ and so on should be determined
from experimental data. The parameters $r$, $s$ and $t$ are
introduced since the both $\alpha_\parallel$ and $V$ works as
the gauge fields of local $U(3)_V$.
\subsection{Higher order corrections}
Higher order terms are suppressed in inverse powers of chiral
symmetry breaking scale and/or heavy quark mass,
so they are less important as far as we
treat low energy light hadrons.
Some examples of higher order terms are shown below;
\beq
      \frac{1}{M}{\rm Tr}[\bar HH{\alpha_\perp}^2], \quad
      \frac{1}{\sqrt{M}}{\rm tr}[\bar TH\psi],\ \dots\  .
\label{eqHOC}
\eeq
For example, the first term can contribute
to $B\rightarrow$ multi $\pi$ decay,
but suppressed by (pion momentum)/$M_B$ in comparison with the
Tr $[\bar HH\alpha_\parallel]$, so it is negligible.
\subsection{Incorporation of symmetry breaking}
The chiral symmetry and the heavy quark symmetry are not exact
because the $u$, $d$ and $s$ quarks have non-zero masses and
neither $m_c$ nor $m_b$ is infinity.
We need corrections of
symmetry breaking in comparing our predictions
with the experimental data.

The light $SU(3)$ flavor symmetry breaking originates from
the current quark masses. The differences among current quark masses
gives rise to the $SU(3)$ breaking in
constituent masses, pseudoscalar meson decay constants, vector meson
coupling constants and hadron mass spectrum.
The  $SU(3)$ breaking corrections are incorporated
by taking the  experimental values for light pseudoscalar
decay constants, light vector meson coupling
constants\cite{EbRe,Kuri}:
\beq
 f \rightarrow f_{ij}, \qquad g_V \rightarrow (g_V)_{ij},
\eeq
and adding $SU(3)$ breaking mass terms of hadrons.
There are two kind of $SU(3)$ breaking mass terms.
One is made from the current
mass matrix of light quarks, $m^0={\rm diag}(m^0_1,m^0_1,m^0_3)$,
the other from the constituent mass matrix,
$m={\rm diag}(m_1,m_1,m_3)$,
where we have assumed that light flavor $SU(2)$ is exact.
Mass terms can be constructed by making invariants with these
mass matrices and hadron fields by assuming that the mass matrices
transform as
\begin{eqnarray}
    m^0 &\rightarrow& z_Lm^0 z_R^{\dag},\\
    m   &\rightarrow& z_V m z_V^{\dag}.
\end{eqnarray}
The transformation property of the former can be understood by
looking at the current quark mass terms,
$\bar q_L m^0 q_R +\bar q_R m^{0\dag} q_L$, while
that of the latter explained in Ref. \cite{EbRe,Kuri}.

Let us see the chiral Lagrangian of heavy mesons
to see how the heavy quark symmetry and the symmetry breaking terms
arise in the heavy sector.The Lagrangian is given to one derivative
on the Nambu-Goldstone bosons as \cite{BurDon,Yans}
\begin{eqnarray}
{\cal{L}}_0
      &=& \sum_{k=D,B} [
	D_\mu\phi_k(D^\mu\phi_k)^{\dag}
	- \phi_k M_k^2\phi_k^{\dag}
	-\frac{1}{2}{F_k^{\mu\nu}} F_{k\mu\nu}^{\dag}
	+ {\phi^*}_{k\mu}{M_{Vk}}^2{{\phi^*}_k^{\mu} }^{\dag}
	\nonumber \\
      & & +\kappa_k\phi_k\alpha_{\perp\mu}{{\phi^*_k}^\mu}^{\dag}
	+\frac{1}{2}g_k\epsilon_{\mu\nu\rho\sigma}
	 F^{\mu\nu}_k\alpha_\perp^\rho {{\phi^*_k}^\sigma}^{\dag}
	 +({\rm h.c.})	],
\end{eqnarray}
where
\begin{eqnarray}
\phi_D ^{(*)}&=& (D^0, D^+, D_s^+) ^{(*)},\qquad
	\phi_B ^{(*)}=(B^-, \bar B^0, \bar B_s^0) ^{(*)}, \\
D_\mu \phi^{(*)}&=& \partial_\mu\phi^{(*)}  +
	i\phi ^{(*)}[r~\alpha_\parallel+(1-r) V]_\mu,\\
F_{\mu\nu}&=& D_\mu {\phi^*}_\nu-D_\nu {\phi^*}_\mu, \\
M_{(V)k}^2 &=& {\rm diag}(M_{(V)k1}^2, M_{(V)k1}^2,M_{(V)k3}^2).
\end{eqnarray}
Note that $\phi^*$  means not the complex conjugate of $\phi$ but
the vector meson. It is assumed that the couplings $\kappa_k$
and $g_k$ do not have light $SU(3)$ flavor dependence.
With the substitution of fields,
\begin{eqnarray}
\phi_{ki}&\rightarrow &
	\frac{1}{\sqrt{2M_{k1}}}\exp [-iM_{k1} v\cdot x]P_{ki},
						\nonumber\\
\phi^{*\mu}_{ki}&\rightarrow &
	\frac{1}{\sqrt{2M_{k1}}}\exp [-iM_{k1} v\cdot x]
						{P^{*\mu}}_{ki}
\qquad (v\cdot P^*=0),\label{hdef}
\end{eqnarray}
this Lagrangian becomes as follows
neglecting $O(\partial ^2/M)$ terms ,
\begin{eqnarray}
{\cal{L}}_0 &=&
	-\frac{1}{2} {\rm Tr}[i\bar H v\cdot DH ] \nonumber\\
      & &+\frac{1}{8}\left( \frac{\kappa_D}{2M_D}
      		+\frac{\kappa_B}{2M_B}\right)
                       {\rm Tr}[
                  \bar H H \gamma_\mu\gamma_5 \alpha_\perp^\mu+
                  \bar H \gamma_\mu\gamma_5 H \alpha_\perp^\mu]
                      				\nonumber\\
      & &+\frac{1}{8}\left( \frac{\kappa_D}{2M_D}
      		-\frac{\kappa_B}{2M_B}\right)
                       {\rm Tr}[
            \bar H\tau_3 H \gamma_\mu\gamma_5 \alpha_\perp^\mu+
            \bar H \tau_3   \gamma_\mu\gamma_5 H \alpha_\perp^\mu]
                      				\nonumber\\
      & &+\frac{1}{8} (g_D +g_B)
                       {\rm Tr}[
                  \bar H H \gamma_\mu\gamma_5 \alpha_\perp^\mu -
                  \bar H \gamma_\mu\gamma_5 H \alpha_\perp^\mu]
                      				\nonumber\\
      & &+\frac{1}{8} (g_D -g_B)
                       {\rm Tr}[
              \bar H\tau_3 H \gamma_\mu\gamma_5 \alpha_\perp^\mu -
              \bar H \tau_3   \gamma_\mu\gamma_5 H \alpha_\perp^\mu]
                      				\nonumber\\
      & & + {\cal{L}}_{RM}, \label{hfl}
\end{eqnarray}
where the Pauli matrix $\tau_3$ works on the heavy
flavor $SU(2)$ indices. The ${\cal{L}}_{RM}$  part represents
residual mass terms;
\begin{eqnarray}
	{\cal{L}}_{RM} &=& \sum_{k=D,B} \left[
	-\frac{(M_{k3}^2-M_{k1}^2)}{2M_{k1}} P_{k3}^{\dag} P_{k3}
	+\frac{(M_{Vk3}^2-M_{k1}^2)}{2M_{k1}}
           		P_{k3}^{*\dag}\cdot P_{k3}^* \right.
						\nonumber \\
           & &   \qquad\quad \left.
           	+\frac{(M_{Vk1}^2-M_{k1}^2)}{2M_{k1}}
           	 ( P_{k1}^{*\dag}\cdot P_{k1}^*+
           	 	P_{k2}^{*\dag}\cdot P_{k2}^*)
           	 	\right]		\nonumber \\
           &=&  -\frac{1}{32} \{
           	\Delta _D+(\theta_D-\theta_{VD})m_1+
           	    \Delta _B+(\theta_B-\theta_{VB})m_1\}
           	    {\rm Tr}[\bar H \sigma_{\mu\nu} H
           	    		\sigma^{\mu\nu}]
           	                   			\nonumber \\
            & &  -\frac{1}{32} \{\Delta _D+
            	             (\theta_D-\theta_{VD})m_1-
           	              \Delta _B-(\theta_B-\theta_{VB})m_1\}
           	       {\rm Tr}[\bar H \tau_3\sigma_{\mu\nu} H
           	       			\sigma^{\mu\nu}]
           	       				\nonumber\\
           & &  +\frac{1}{32} \{(\theta_D-\theta_{VD})
           				+(\theta_B-\theta_{VB})\}
           	       {\rm Tr}[\bar H \sigma_{\mu\nu} H
           	       				\sigma^{\mu\nu}m]
           	       				\nonumber \\
           & &  -\frac{1}{32} \{(\theta_D-\theta_{VD})
           		-(\theta_B-\theta_{VB})\}
           	       {\rm Tr}[\bar H \tau_3\sigma_{\mu\nu} H
           	       				\sigma^{\mu\nu}m]
           	       				\nonumber\\
           & &  +\frac{1}{16} \{3\Delta _D-
           		(\theta_D-3\theta_{VD})m_1+
           	         3\Delta _B-(\theta_B-3\theta_{VB})m_1\}
           	                   {\rm Tr}[\bar H H]
           	                   		\nonumber \\
           & &  +\frac{1}{16} \{3\Delta _D-
           		(\theta_D-3\theta_{VD})m_1-
           	         3\Delta _B+(\theta_B-3\theta_{VB})m_1\}
           	                   {\rm Tr}[\bar H \tau_3 H]
           	                   		\nonumber\\
          & &  +\frac{1}{16} \{(\theta_D-3\theta_{VD})
          			+(\theta_B-3\theta_{VB})\}
           	       		{\rm Tr}[\bar H H m]
           	       				\nonumber \\
          & &  +\frac{1}{16} \{(\theta_D-3\theta_{VD})
          		-(\theta_B-3\theta_{VB})\}
           	       {\rm Tr}[\bar H \tau_3 Hm] \nonumber \\
          & &  + O\left(\frac{\Delta^2}{M}, \frac{m^2}{M} \right) ,
 \end{eqnarray}
where the parameters are defined as follows;
\begin{eqnarray}
\Delta_k &\equiv & M_{Vk1}-M_{k1}, \\
\theta_k (m_3-m_1) &\equiv & M_{k3}-M_{k1},\\
\theta_{Vk} (m_3-m_1) &\equiv & M_{Vk3}-M_{Vk1}.
\end{eqnarray}
The above equations show that the following conditions should
hold for the symmetries
to be exact:\\
heavy $SU(2)$ flavor symmetry
\beq
\frac{\kappa_D}{2M_D}=\frac{\kappa_B}{2M_B},
\quad g_D=g_B, \quad \Delta_D=\Delta_B,
\quad \theta_D=\theta_B,\quad \theta_{VD}=\theta_{VB}.
\eeq
heavy spin symmetry
\beq
\frac{\kappa_D}{2M_D}=g_D,\quad \frac{\kappa_B}{2M_B}=g_B,
\quad \Delta_D=\Delta_B=0,
\quad \theta_D=\theta_{VD},\quad \theta_{B}=\theta_{VB}.
\eeq
Therefore, the following conditions are necessary for the
Lagrangian ${\cal{L}}_0$ to be equivalent to the effective
heavy field Lagrangian  ${\cal{L}}_H$;
\begin{eqnarray}
g_D &=&g_B= \frac{f_D}{2M_D}=\frac{f_B}{2M_B}\equiv \lambda\\
\Delta_D&=&\Delta_B=0,\\
\theta_D&=&\theta_{VD}=\theta_B=\theta_{VB}.
\end{eqnarray}
The first condition is derived by using the interpolating fields also
in the work by Yan et.al. \cite{Yans}. The deviation from the
above conditions gives the symmetry breaking terms which are not
considered in the lowest order analysis except the breaking in mass.
The reason why the mass breaking
should be taken into account is given as follows:
When we work in the effective fields, the denominator of the
propagator of the field $X$ is given by $v\cdot (p-Mv)-\delta$ with
the breaking mass terms taken into account,
where $p$ is the momentum of the propagating field , $M$ is
either $M_D$ or $M_B$ depending the heavy flavor of the field,
and $\delta=M_X-M$. The subtraction $Mv$
appears because the effective field is defined as in Eq.(\ref{hdef}).
Let us express the momentum as $p=M_X v +k$. Then we have
\beq
  v\cdot(p-Mv)-\delta=v\cdot(M_X v+k-Mv)-\delta =v\cdot k.
\eeq
The energy scale of $v\cdot k$ is of the order of the QCD scale,
and it is not far larger than the magnitude of $\delta$ for the
heavy vector mesons and the heavy mesons with strange flavor.
Therefore, the breaking effect in mass cannot be neglected
in the propagator of heavy fields.

The symmetry breaking terms for heavy baryons can be introduced
in the similar way. The spin symmetry breaking appears in the
axial vector coupling between heavy baryons,
\beq
{\rm tr}[\bar T \gamma\gamma_5 T \cdot \alpha_\perp] .
\eeq
But we do not consider this breaking in the lowest order analysis
as in the case of heavy mesons.

In the analysis given later we consider symmetry breaking effect
in the light pseudo scalar decay constants, light vector meson
coupling constants and hadron masses, but the correction in heavy
hadron coupling constants are neglected.

\section{Incorporation of weak interaction}
\label{sec:weak}

At low energy scale far below the W boson mass
the weak interaction is described by the effective Hamiltonians
expressed in terms of 4-Fermi operators of quarks and/or leptons
which are obtained by integrating out $W$, $Z$ bosons and top quark.
Effective weak interaction Hamiltonian of hadrons should have
the same transformation property under the
symmetry as the quark effective Hamiltonian. It is constructed
by replacing quark currents with the hadron currents of the same
transformation properties under the symmetry.

Left-handed quark current coupled to an external vector
field $L$ is given as follows,
\beq
J_L \cdot L =(\bar q, \bar Q_{v'})
       \left( \begin{array}{cc}
              L_l & L_m \\
              {L_m}^{\dag} & L_h
              \end{array} \right)
\cdot\gamma \left( \frac{1-\gamma_5}{2} \right)
       \left( \begin{array}{c} q \\ Q_v \end{array} \right),
\eeq
where $q$ and $Q$ represent light quark field and effective heavy
quark field, respectively. The above coupling is invariant
if the external field transforms as
\beq
\begin{array}{lcl}
L_l \rightarrow z_L L_l {z_L}^{\dag}, & & \\
L_h\rightarrow z_H^{v'} L_h (z_H^{v})^{\dag} &,\quad &
L_h\cdot \gamma \frac{(1-\gamma_5)}{2} \rightarrow
S^{v'}L_h\cdot \gamma \frac{(1-\gamma_5)}{2}(S^{v})^{\dag}, \\
L_m\rightarrow z_L L_m (z_H^{v})^{\dag} &,\quad &
L_m\cdot \gamma \frac{(1-\gamma_5)}{2} \rightarrow
L_m\cdot \gamma \frac{(1-\gamma_5)}{2}(S^{v})^{\dag}.
\end{array}
\eeq

We construct invariants containing the external filed $L$ in order to
obtain the hadron currents which has the same transformation property
as the quark currents.

The invariants which contain $L_l$ are given in the lowest
order of chiral expansion as
\beq
{\rm tr}[\xi^{\dag} L_l \xi \cdot (\alpha_\parallel -V)],\quad
      {\rm tr}[\xi^{\dag} L_l \xi\cdot \alpha_\perp],\quad
      {\rm tr}[\bar\psi \xi^{\dag} L_l \xi\cdot
      				\gamma (\gamma_5) \psi],\quad
      {\rm tr}[\bar\psi \gamma (\gamma_5) \psi \cdot
      				\xi^{\dag} L_l \xi].
\eeq
The term like ${\rm Tr}[\bar H H v\cdot\xi^{\dag} L_l \xi]$ also
exists, but they are not important in the weak decay of heavy
hadrons which are considered later.

The invariants containing  $L_h$ are to lowest order of chiral
and $1/M$ expansion
\beq
{\rm Tr}[\bar H_{v'} L_h\cdot \gamma\frac{(1-\gamma_5)}{2} H_v],
\quad
{\rm Tr}[\bar T_{v'} L_h\cdot \gamma\frac{(1-\gamma_5)}{2} T_v].
\eeq

The invariants involving $L_m$ is interesting since they lead to
the heavy to light current which is significant in the
determination of $|V_{ub}|$. They are given in lowest order
of chiral and $1/M$ expansion as
\beq
\begin{array}{ll}
{\rm Tr}[\xi^{\dag}L_m\cdot \gamma\frac{(1-\gamma_5)}{2}H_v], & \\
{\rm tr}[\bar \psi \xi^{\dag}
    	L_m\cdot \gamma\frac{(1-\gamma_5)}{2}T_v], &
 \ {\rm tr}[\bar \psi \xi^{\dag} v\cdot\gamma
    	L_m\cdot \gamma\frac{(1-\gamma_5)}{2}T_v],
 \end{array}
\eeq
There are two types of invariants between light baryon and
heavy baryon as given above because we can insert $v\cdot\gamma$ in
front or back of the heavy fields as far as
the invariance is not spoiled. The insertion
of $v\cdot\gamma$ spoils the invariance or keeps the form unchanged
in other invariants.

Now, hadron currents are obtained by removing the external
field $L$ from the above invariants. The light to light current
is obtained as
\begin{eqnarray}
 (J_{lL})_{ij}&=&
      a\ [\xi (\alpha_\parallel-V)_\mu \xi^{\dag}]_{ij} +
      b\ [\xi (\alpha_\perp)_\mu \xi^{\dag}]_{ij}, \nonumber\\
      & &
      +\sum_k [\bar\psi\xi^{\dag}\gamma_\mu (c_1+c_2 \gamma_5)]_{kj}
                        [\xi \psi]_{ik}+
       [\xi \bar\psi\gamma_\mu
       		(c_3+c_4 \gamma_5)\psi \xi^{\dag}]_{ij}
       +\cdots.
 \end{eqnarray}
The coefficients, $a$, $b, \cdots$, are determined
once we fix the symmetric part of
the Lagrangian because this is the Noether current of the light
flavor chiral symmetry.

The heavy to heavy current is given as
\beq
(J_{hL})_{ab} =
        \xi_h {\rm Tr}
         [\bar H_{v'b}\gamma(1-\gamma_5) H_{va}]/2 +
        \eta_h  \bar T_{v'b}\gamma(1-\gamma_5) T_{va}/2+ \cdots .
\eeq
This is not a Noether current in general since the velocities of
two heavy fields are not necessary the same.
The coefficients, $\xi_h$ and $\eta_h$,
are constant at the lowest order,
but should be regarded as the function of $v\cdot v'$.
They are  so-called the Isgur-Wise functions and take a fixed value
when the velocities coincide with each other
because the current becomes Noether current in that case.

The heavy to light current is obtained as
\beq
(J_{mL})_{ia} =f_m {\rm Tr}[(\xi^{\dag})_{ji}
                     \gamma(1-\gamma_5) H_{va}^j]/2+
         (\bar \psi\xi^{\dag})_{ji}(F_1+F_2 v\cdot\gamma)
         \gamma(1-\gamma_5)
          T_{va}^j/2+\cdots.
\eeq
The coefficients, $f_m$, $F_1$ and $F_2$, should be taken as
functions of $v\cdot q$, where $q$ is the momentum of the light
hadron. They are not fixed from the symmetry alone since we do not
have a symmetry which relates heavy and light hadrons.

Right-handed current can be obtained in a similar way.
Scalar and Tensor currents are also obtained by introducing scalar
or tensor external fields.

Weak effective Hamiltonian of hadrons can be constructed by
using the currents given so far. The effective weak Hamiltonians
of semi-leptonic and non-leptonic decays are
given at the tree level of quarks as follows
\begin{eqnarray}
{\cal{H}}_{sl} &=& 2\sqrt{2} G_F V_{ij}
    \bar u_{Lj\alpha} \gamma d_{Lj\alpha}
    	\cdot \bar l_L \gamma \nu_{L} +{\rm (h.c)}, \\
{\cal{H}}_{nl} &=& 2\sqrt{2} G_F V_{ij} V_{kl}^*
    \bar u_{Li\alpha }\gamma d_{Lj\alpha}
    	\cdot \bar d_{Ll\beta} \gamma u_{Lk\beta} +{\rm (h.c)} \\
             &=& 2\sqrt{2} G_F V_{ij} V_{kl}^*
    \bar u_{Li\alpha} \gamma u_{Lk\beta}
    	\cdot \bar d_{Ll\beta} \gamma d_{Lj\alpha} +{\rm (h.c)} ,
\end{eqnarray}
where the Greek letters express color indices.
The last equality is obtained by using Fiertz transformation.
Effective Hamiltonian in hadron form is given by replacing the
quark current with the corresponding hadron current.
When the second form of the non-leptonic decay Hamiltonian is used,
color factor $1/N_C$ should be considered.
Feynman rule calculation of the decay rate can be done with
thus obtained effective weak Hamiltonian and the strong
interaction Lagrangian given in Sec.\ref{sec:lag}.
\section{Applications to weak decays}
\label{sec:apl}

Let us investigate some heavy hadron decays. For simplicity,
we do not consider here QCD correction to the weak Hamiltonian of
quarks, nor hadron loop contribution. We do not treat a decay
where significant contribution comes from an effective Hamiltonian
which is obtained after integrating out a quark loop, like penguin
diagram. These effects are important and interesting in some
processes, but we leave the investigation of them
in our future study.
\subsection{$D^0 \rightarrow K^- e^+ \nu _e$ decay}
\label{sec:DKl}

We analyze $D^0 \rightarrow K^- e^+ \nu_e$ decay
to fit the $D_s$ decay constant and the strong coupling
constant $\lambda$. The effective weak Hamiltonian involving this
process is given as
\beq
 {\cal H}_{eff} = 2\sqrt{2} G_F V_{cs}^*
                   ({\bar s}_L \gamma_\mu c_L)
                   ({\bar \nu}_{eL} \gamma^\mu e_L)
                  + \rm{(h.c.)}.
\eeq
The quark current $\bar s_L \gamma_\mu c_L$ is converted to hadron
current.
\beq
   f_{m0}{\rm Tr}
          \left[ (\xi^{\dag})_{j3} \gamma_\mu {{1-\gamma_5} \over 2}
          				H_c^j \right]
           +
    f_{m1}{\rm Tr}
          \left[ (\alpha_\perp^\nu \xi^{\dag})_{j3}
          \gamma_\nu\gamma_\mu
          {{1-\gamma_5} \over 2} H_c^j \right]+\cdots.
\eeq
We find that the unknown factor $f_{m0}$ is related to
the decay constant of $D_s$,
\beq
 f_{m0} = - {i \over 2} f_{Ds} \sqrt{M_{Ds}},
\eeq
by looking at the lowest order term of light pseudoscalar expansion
of the current and the definition of the decay constant.
\beq
\langle 0 | {\bar s}_L \gamma_\mu c_L | D_s(p) \rangle
 = i {{f_{Ds}} \over {\sqrt{2}}} p_\mu .
\eeq
(Note that our normalization of decay constant
gives $f_\pi = 93$ MeV.) The Feynman diagram for this decay process
is shown in Figs.~\ref{fig1} (a) and (b).
The form factor is obtained up to the $f_{m1}$ term as
\beq
 f_\pm (v \cdot p_K) =
        - {1 \over 2} \sqrt{\frac{M_{Ds}}{M_D}} {f_{Ds} \over  f_K}
   \left[ (1-\lambda) \pm { {\lambda M_{{Ds}^*} } \over
     {v \cdot p_K+\Delta} } \right],
\eeq
where $\Delta=M_{{Ds}^*}-M_D$.
The form factor has one pole structure if the coupling
$\lambda$ is unity, although the pole type Feynman diagram does
not necessary dominate for $\lambda =1$.
The form factor $f_-$ does not contribute to the semileptonic
decay when the lepton mass is neglected.
The energy spectrum is calculated as follows,
\beq
 {{d\Gamma} \over {dq^2}} = {{G_F^2 |V_{cs}|^2} \over {192\pi^3}}
 		M_D^3  |f_+(v\cdot p_K)|^2 \left[
             {{((M_D+m_K)^2-q^2) ((M_D-m_K)^2-q^2)} \over {M_D^4}}
                                           \right]^{3/2},
\eeq
where $q^2=M_D^2+m_K^2-2M_D v\cdot p_K$.
We have compared our result of energy spectrum with the Mark III
data\cite{MarkIII}.
The result is shown in Fig.\ref{fig2}, where the prediction from the
one pole fit of the form factor taken in Ref.~\cite{MarkIII} is
also shown for comparison.
They took $f_+(q^2)=f_+(0)m^2/(m^2-q^2)$ with
$m=1.8$ GeV and $|f_+(0)/V_{cs}|=0.72$.
While our prediction gives for $\lambda=1$
\beq
f_+ = -\left( \frac{\sqrt{M_{Ds}M_D}f_{Ds}}{M_{Ds*}f_K}\right)
	\frac{M_{Ds*}^2}{M_{Ds*}^2-q^2+m_K^2-\Delta^2},
\eeq
which gives the Figure \ref{fig2} (a)
for $f_{Ds}$=90, 100 and 110 MeV.

We should be careful in comparing our result with the
experimental data. There is a region where our results cannot
taken to be serious because we
have used chiral expansion for light mesons.
Chiral expansion will not give a good approximation when the
momentum of Nambu-Goldstone boson becomes large.
As long as we require
the energy of $K$ meson to be less than some hadronic scale,
say $\rho$ meson
mass, then our result will not give a good prediction
for $q^2 \leq 0.8\ {\rm GeV}^2$ or so.
Bearing this point in mind we check whether the 3 points
($q^2$=1.0, 1.4, 1.8 ${\rm GeV}^2$) of experimental data agree
with our predictions.
If the one pole structure of the form factor ($\lambda=1.0$) is
assumed, the experimental data favors $f_{D_s} \simeq 100$ MeV.
If we take $f_{D_s} = 160$ MeV suggested
by the recent measurement\cite{WA75},
the experimental data favors $\lambda \sim 0.4$,
which indicates the deviation from one pole structure of
the form factor.

We have investigated the effects of higher order term
in chiral expansion by
considering $f_{m1}$ term. Predictions are made
for $f_r\equiv \sqrt{2M_D}f_{m1}/f_{Ds}=-0.1$ to $0.1$
with $\lambda=0.4$ and $f_{Ds}=160$ MeV.
The result is shown in Fig.~\ref{fig3},
which shows that inclusion of this higher order term
does not drastically change the curve of spectrum, and plays
similar role as changing the value of $\lambda$.

The coupling constant $\lambda$ can be fixed also by estimating
the process $D^{*+} \rightarrow D^0 \pi^+$. Our approximation
is good there since the momentum of $\pi$ is small in this
2-body decay. The width is obtained as
\begin{equation}
 \Gamma={{\lambda^2} \over {96\pi}} {{M_D^2} \over {f_\pi^2}} M_{D^*}
        \left[
        {{ ((M_{D^*}+m_\pi)^2-M_D^2) ((M_{D^*}-m_\pi)^2-M_D^2) }
                \over {M_{D^*}^4}}
        \right]^{3/2}.
\end{equation}
The $M_{D*}-M_D$ mass difference and velocity difference
 are considered in the phase space integration.
The experimental upper
bound $\Gamma<1.1$MeV\cite{PD} gives $\lambda<2.6$.
If we use the recent measurement for branching ratio
 ${\rm Br}(D^{*+} \rightarrow D^0 \pi^+) = 68.1 \pm 1.0 \pm 1.3$
 \%\cite{CLEO}
 and the theoretical prediction for total width $\Gamma = 59.4$ keV
 ( Model(b) in \cite{Kamal-Xi}),
 we get $\lambda \simeq 0.49$.
This also indicates the deviation from
the one pole structure of the form factor.
\subsection{$ \Lambda_b \rightarrow p l \bar\nu$ decay}

The $\Lambda_b \rightarrow p l \bar\nu$ decay mode is interesting
because it leads to the possibility of
model independent determination of $|V_{ub}|$.
The effective Hamiltonian in quarks and leptons is given as
\beq
{\cal{H}}_{eff}=2\sqrt{2}G_F V_{ub} \bar u_L\gamma_\mu b_L
	\bar l_L\gamma^\mu \nu_L + {\rm (h.c.)} .
\eeq
The quark current is translated to hadron currenxt
containing $\Lambda_b $ and proton.
\beq
\bar u_L\gamma_\mu b_L \rightarrow
      \bar p (F_1+F_2 v\cdot \gamma)\gamma_\mu
      	\Lambda_{bL}.
\eeq
The Feynman diagram of lowest order contribution is shown in
Fig.~\ref{fig4}. This decay is described by two
parameters, $F_1$ and $F_2$, as first shown by
Mannel, Roberts and Ryzak\cite{MRR} by using tensor
method\cite{Georgi3}. (There is also $B^* $ pole contribution
as shown in Fig.~\ref{fig5}
if we consider the higher order term given in Eq.~(\ref{eqHOC}).
But we do not consider it here.)
These parameters are constant at the lowest level of the
effective Lagrangian, but can be regarded as the form factors
which are the functions of $v \cdot p_p$ ,
where $v$ is the velocity of $\Lambda_b$ and $p_p$ is the proton
momentum. The differential decay rate is calculated as
\begin{eqnarray}
\frac{{\rm d}\Gamma (\Lambda_b \rightarrow pl\bar\nu)}{dE_p dE_l}
   & =& \frac{G_F^2|V_{ub}|^2}{\pi^3}
     [(F_1^2-F_2^2)\{M_{\Lambda b}(E_p+E_l) -
     \frac{1}{2}(M_{\Lambda b}^2+m_p^2)\} 	\nonumber\\
     & &\qquad +2m_pE_l F_1F_2+2E_pE_l F_2^2]
     				(M_{\Lambda b}-E_p-E_l).
\end{eqnarray}
Similar decay rate is calculated also
for $\Lambda_c^+ \rightarrow \Lambda \bar l \nu$ decay.
The ratio of these two becomes at the stopping $p$, $\Lambda$ limit.
\beq
 \frac{
    \left.
    {\rm d}\Gamma (\Lambda_b \rightarrow pl\bar\nu)/{dE_p dE_l}
    \right|_{{\rm stop}\ p}
    	 }{
    \left.
    {{\rm d}\Gamma
    (\Lambda_c \rightarrow \Lambda \bar l\nu)}/{dE_\Lambda dE_l}
    \right|_{{\rm stop}\ \Lambda}
    	 }
              =  \frac{3}{2}
                 \frac{|V_{ub}|^2}{|V_{cs}|^2}
                 \frac{m_p (M_{\Lambda b}-m_p)^2}
                     {m_\Lambda (M_{\Lambda c}-m_\Lambda)^2}
                \frac{|F_1(m_p) + F_2 (m_p)|^2}
                {|F_1(m_\Lambda) + F_2 (m_\Lambda)|^2}.
\eeq
The factor $3/2$ comes from light flavor $SU(3)$ relation.
The ratio of form
factors becomes unity in the $SU(3)$ symmetric limit.
So we can obtain $|V_{ub}|$
from the measurement of this ratio of decay rates without hadron
model dependence. Still we have to take QCD
correction, $SU(3)$ breaking effect and 1/M correction into
account to obtain more precise value of $|V_{ub}|$.
If $\Lambda_c \rightarrow n \bar l \nu$ decay can
be measured, the $SU(3)$ breaking effect is absent.
\beq
   \frac{
    \left.
    {{\rm d}\Gamma (\Lambda_b \rightarrow pl\bar\nu)
                }/{dE_p dE_l}
    \right|_{{\rm stop}\ p}
      }{
    \left.
    {{\rm d}\Gamma (\Lambda_c \rightarrow n \bar l\nu)}/
    			                    {dE_n dE_l}
    \right|_{{\rm stop}\ n}
      }
	=
           \frac{|V_{ub}|^2}{|V_{cd}|^2}
           \frac{ (M_{\Lambda b}-m_N)^2} { (M_{\Lambda c}-m_N)^2}.
\eeq
\section{Concluding remarks}
\label{sec:conc}

We have constructed a Lagrangian of heavy and light hadrons
from $\pi$ to $\Lambda_b$ by utilizing the chiral symmetry for
light flavors and the heavy quark symmetry.
Though only $S$ wave bound state hadrons of quarks are discussed
here, $P$ or higher wave states can also be incorporated as far as
their transformation property under the symmetry is given.
We can reason possible form of higher order correction and
symmetry breaking effect from the symmetry property of hadrons,
and examine if it is significant or not by fitting
the prediction with the experimental data.

We have analyzed $D^0 \rightarrow K^- e^+\nu_e$ decay and
$\Lambda_b \rightarrow p l \bar \nu$ decay as the examples of
application of our formalism. We have fitted the $D_s$ decay
constant and the coupling constant among heavy meson,
heavy vector meson and light meson by
comparing the $q^2$ spectrum of our prediction with
the Mark III data of semileptonic $D$ decay. We have pointed out
the possibility of model independent determination
of $|V_{ub}|$ in $\Lambda_b \rightarrow$ nucleon semileptonic decay.

Though only semi-leptonic decays are discussed in the application,
non-leptonic decay can also be handled with our formalism.
For two body decays of heavy hadrons the
resulting light hadron in general has a high momentum. So the chiral
perturbation may not be good enough. But if the factorization
anzats works well, our formalism will be available also for
two body decays. As a matter of course our formalism is available
for multi body decay where light hadrons have low momentum.

Tree level quark diagram is dominated in the processes discussed
here. There are some processes where quark loop diagram
contribution is significant, like penguin contribution
in $B \rightarrow \pi \pi $ decay. To deal with such a process
quark effective Hamiltonian is constructed
first by integrating out quark loop, and then translate the
quark Hamiltonian into hadronic one which has
the same transformation property under the symmetry.

Our discussion has been almost limited to the lowest order
contributions in chiral expansion and $1/M$ expansion.
We need take higher order,  symmetry breaking and QCD corrections
into account to make our prediction more precise.
This will be done in our subsequent paper.

\acknowledgments

We would like to thank Dr. M. Tanaka of KEK theory group
for fruitful discussions and comments,
and Professor A. I. Sanda for encouragement.
T. K.'s work is supported in part by Grant-in Aids for Scientific
Research from the Ministry of Education, Science and Culture
(No.03302015 and 02640230).
\figure{Feynman diagrams contributing
             to  $D^0 \rightarrow K^- e^+ \nu_e$ decay.
             The solid square represents weak interaction
             while the solid circle strong interaction:
             (a) contact term contribution;
             (b) $D_s^*$ pole contribution.
             \label{fig1}}
\figure{
Energy spectrum of $D^0 \rightarrow K^- e^+ \nu_e$ decay.
The cross lines show MarkIII data. The dashed line is the result of
one pole fit of the form factor given in Ref.~\cite{MarkIII}:
(a)  $f_{D_s}=90,\ 100,\ 110 $MeV with $\lambda_1=1.0$;
(b) $\lambda =0.2,\ 0.3,\ 0.4,\ 0.5$ with $f_{D_s}=160$MeV.
\label{fig2}
}
\figure{Same as Fig. 2 but a higher order term considered:
	$f_{D_s}=160$MeV, $\lambda_1=0.4$ and
	$f_r=-0.1,\ 0.0,\ 0.1$.
	\label{fig3}}
\figure{
	Lowest order contribution
	to $\Lambda_b  \rightarrow p l \bar\nu$ decay .
\label{fig4}}
\figure{
	Contribution to $\Lambda_b \rightarrow p l \bar\nu$ decay
	from a higher order term.
\label{fig5}}
\begin{table}
\caption{Transformation properties of hadrons under the symmetry.}
 \begin{tabular}{ccccc}
    field & & $SU(3)_V$ & $SU(2)_{heavy}$ & spin sym. \\
     \tableline
   $\alpha_\perp $& $\rightarrow$ &
   $z_V\alpha_\perp {z_V}^{\dag}$ & & \\
   $\alpha_\parallel -V $& $\rightarrow$ &
      $z_V(\alpha_\parallel -V){z_V}^{\dag}$ & & \\
   $\psi$ & $\rightarrow$ & $z_V\psi {z_V}^{\dag} $& & \\
   $H_v$ & $\rightarrow$ & $H_v{z_V}^{\dag}$ &
   $z_H^v H_v $& $S^v H_v$ \\
   $T_v$ & $\rightarrow$ & $T_v{z_V}^{\dag}$ &
   $z_H^v H_v$ & $S^v T_v$
  \end{tabular}
\label{table1}
\end{table}
\end{document}